\documentstyle[preprint,aps]{revtex}

\newtheorem{theorem}{Theorem}
\newtheorem{acknowledgement}[theorem]{Acknowledgement}

\begin{document}
\title{A model Hamiltonian for $MgB_{2}$\ which takes into account its unusual
phononic features.}
\author{M. Acquarone$^{1,2}$ and L. Roman\'{o}$^{2}$}
\address{$^{1}$IMEM-CNR, Parco Area delle Scienze, 43100 Parma (Italy); $^{2\text{ }}$%
Unit\`{a} INFM e Dipartimento di Fisica, Universit\`{a} di Parma, Parco Area%
\\
delle Scienze, 43100 Parma (Italy).}
\date{\today}
\maketitle
\pacs{74.20-z, 74.70,-b}

\begin{abstract}
Taking as a starting point the results of LDA calculations, which show that
in $MgB_{2}$\ the phonons have a strong quartic anharmonicity and that the
bond-stretching electron-phonon interaction (EPI) has both a linear and a
large quadratic component, we propose a model Hamiltonian which succesfully
matches a number of experimental evidences. We relate the single critical
temperature for both superconducting gaps to a phonon-induced inter-band
coupling whose amplitude increases with temperature. We also obtain phonon
frequencies and linewidths depending on the band filling, as well as band
energies and hybridization amplitudes depending on the phonon number.

PACS:74.20-z, 74.70-b
\end{abstract}

\section{Introduction.}

\bigskip The electronic structure of the $40$ K superconductor $MgB_{2}$ is
characterized by the presence at the Fermi level of two hybrid bands ($%
\sigma $ and $\pi )$ of very different character\cite{mazin}. This feature
reflects itself in the experimental evidence of two different gaps\cite
{bouquet,giubileo,schmidt,iavarone,tsuda}, which however, in the absence of
magnetic fields, have a common critical temperature\cite{choi-nature}\label%
{mazin2}. The observation of a large Boron isotope effect\cite{budko} rules
out the applicability of theories of Coulomb-interaction-driven two-band
superconductivity\cite{kondo}, suggesting instead that the
electron-phonon-interaction(EPI) \ is the key factor.

According to the standard theory of the EPI-driven two-band
superconductivity \cite{suhl}, a single critical temperature for both gaps
implies an interaction between the bands contributing to the Fermi surface.
The microscopic origin of this interaction for the $\sigma $ and $\pi $
bands in $MgB_{2}$ is not yet clarified. Impurity scattering can be ruled
out \cite{mazin2,jemina} and, to the best of our knowledge, no other precise
suggestion has been advanced for the EPI\ scenario. An estimate of the
interband coupling strength has been given in Ref.\cite{dahm}, based on the
band structure calculation of Ref.\cite{liu}, yielding a small, but
decidedly non-negligible value. The strong temperature dependence of the
anisotropy of the critical field\cite{dahm} indicates that the interband
coupling increases with temperature. The present work suggests that such
coupling might be due to the unusual phononic structure of $MgB_{2}.$ There
is a general consensus that the dominant electron- phonon interaction is due
to a modulation of the inter-site hopping amplitudes due to the
bond-stretching vibration of the Boron ions \cite
{liu,dolgov,Yildirim,kunc,choi,postorino}. By working out the corresponding
Eliashberg's $\lambda $ , Ref.\cite{dolgov} shows that there is a good
agreement with the results following the LDA data, and with the experiments.
In one-dimensional materials, this type of interaction is usually termed the
Su-Schrieffer- Heeger (SSH) interaction, and we' ll use this terminology
also in the present context for conveniency.

The unusual phononic features (first revealed by ab-initio calculations of
the electron and phonon band structures \cite{liu,Yildirim,kunc,choi,bohnen}%
) are the presence of anharmonic contributions (up to fourth order in the
displacements) to the phononic Hamiltonian, and of \ both a linear and a
quadratic term in the SSH\ interaction. Experimental evidence of
anharmonicity comes from neutron \cite{Yildirim} and Raman scattering \cite
{renker} data. More specifically, the first-principle calculations in Ref. 
\cite{bohnen} find that the $E_{2g}$ branch has, along the $\Gamma -A$ line
in the Brillouin zone, an energy around 120 meV in the harmonic compound $%
AlB_{2}$ and of only 70 meV in $MgB_{2}$, in good quantitative agreement
with the neutron scattering data\cite{Yildirim}. Additionally, Ref. \cite
{renker} presents first-principle calculations of the evolution of the
phonon spectra when $Al$ substitutes $Mg$ agreeing with Raman data, which
confirm the frequency softening on passing from harmonic $AlB_{2}$ to
anharmonic $MgB_{2}$. One important aspect of such measurements is that, as $%
Al$ substitution changes the occupancy of the bands at the Fermi surface,
those data show that the phonon frequency and lifetime both depend on the
band filling. So, in relating those data to the change in phononic
properties, one should be able to disentangle the effects of anharmonicity
from those of band-filling variations. However, one must also mention that
Raman and infrared data of Ref.\cite{postorino} for $Mg_{1-x}Al_{x}B_{2}$
when $0\leq x\leq 0.50$ , on the contrary, find the frequency of the $E_{2g}$
mode almost insensitive to the $Al$ content.

Another theoretically predicted effect of the anharmonicity is the reduction
of the averaged electron-phonon coupling, as expressed by Eliashberg's $%
\lambda $. For that, the experimental evidence is not so clear. Indeed, the
first-principle calculation of the phonon spectra of Ref. \cite{choi} are in
excellent agreement with the experimental data, and they yield at the same
time that the strength of $\lambda $ is reduced in the anharmonic case.
However, if one takes the phonon linewidth as roughly expressing the
combined intensity of the electron-phonon interactions, both Ref.\cite
{postorino} and \cite{renker} show that it strongly decreases with $x.$ As
Eliashberg's $\lambda $ for a bond-stretching interaction\cite{barisic}
depends on structure, band-filling and frequencies, it is difficult to
precisely identify the cause(s) of the observed effects.

The ab initio numerical calculations\cite{liu,Yildirim,kunc,choi,bohnen}\
have yielded valuable insights about the electronic and phononic structure
of $MgB_{2}$, which we take as the input information for the work presented
here. Our aim in this paper is to propose a model Hamiltonian which
represents the physics implied by the results of the first-principle
calculations as far as the phononic features are concerned. We have no
ambition of giving detailed quantitative results. However, our model is
quantitatively consistent with the numerical results of Refs.\cite
{liu,Yildirim,choi}. While suggesting a plausible mechanism for the
inter-band coupling, and therefore justifying\cite{suhl} the presence of a
single critical temperature for both superconducting gaps \cite
{bouquet,giubileo,schmidt,iavarone,tsuda}, at the same time it also
qualitatively allows for other experimentally observed features: the
increase with temperature of the inter-band coupling \cite{dahm} and the
fact that the phonon frequency and linewidth both depend on band-filling\cite
{postorino,renker}. In particular, we find indications that the frequency
hardening on $Al$\ substitution can not be accounted for by the
anharmonic-to-harmonic change only.

Detailed quantitative estimates based on the proposed model are left for
future work.

\section{The electronic Hamiltonian.}

\bigskip Our model of the electronic structure of $MgB_{2}$ by a Hamiltonian
has two bands, labelled $c$ and $d,$ which hybridize through an inter-site
hopping term. Then, in the real space representation, the bare electronic
Hamiltonian reads: 
\begin{equation}
H_{el}^{{}}=\varepsilon ^{c}\sum_{l\sigma }n_{l\sigma }^{c}+\varepsilon
^{d}\sum_{l\sigma }n_{l\sigma }^{d}+\sum_{l\langle j\rangle \sigma }\left[
t_{lj}^{cc}c_{l\sigma }^{\dagger }c_{j\sigma }^{{}}+t_{lj}^{dd}d_{l\sigma
}^{\dagger }d_{j\sigma }^{{}}\right] +\sum_{l\langle j\rangle \sigma
}t_{lj}^{cd}\left( c_{l\sigma }^{\dagger }d_{j\sigma }^{{}}+d_{j\sigma
}^{\dagger }c_{l\sigma }^{{}}\right)  \label{eq.Hel}
\end{equation}

where $\sum_{l\langle j\rangle }$ means summing on the $z$ nearest neigbors $%
j$ of a given site $l$, and then on $l.$ In $MgB_{2}$ one expects that $%
t_{lj}^{cc},t_{lj}^{dd}\gg t_{lj}^{cd}$ \cite{mazin,dahm,dolgov}. The
electron-phonon coupling parameters in the SSH scenario are derivatives of
the hopping amplitudes with respect to the inter-site distance. By using $%
c_{l\sigma }^{\dagger }=N^{-1/2}\sum_{k}c_{k\sigma }^{\dagger }\exp \left(
ikR_{l}^{{}}\right) $ we pass to the reciprocal space representation,
yielding : 
\begin{equation}
H_{el}^{{}}=\sum_{k\sigma }\left( \varepsilon ^{c}+zt_{k}^{cc}\right)
n_{k\sigma }^{c}+\sum_{k\sigma }\left( \varepsilon ^{d}+zt_{k}^{dd}\right)
n_{k\sigma }^{d}+\sum_{k\sigma }zt_{k}^{cd}\left( c_{k\sigma }^{\dagger
}d_{k\sigma }^{{}}+d_{k\sigma }^{\dagger }c_{k\sigma }^{{}}\right)
\end{equation}
where $t_{k}^{\mu \nu }=z^{-1}\sum_{\langle j\rangle }t_{lj}^{\mu \nu }\exp %
\left[ ik\left( R_{l}^{{}}-R_{j}^{{}}\right) \right] $ \ with $\mu ,\nu =c,d$%
. To diagonalize $H_{el}^{{}}$ we express the bare operators \ $c_{k\sigma
}^{\dagger },d_{k\sigma }^{\dagger }$\ through the hybridized operators $%
\alpha _{k\sigma }^{\dagger },\beta _{k\sigma }^{\dagger }$ according to: 
\begin{equation}
c_{k\sigma }^{\dagger }=\alpha _{k\sigma }^{\dagger }\cos \varphi
_{k}^{{}}+\beta _{k\sigma }^{\dagger }\sin \varphi _{k}^{{}}\qquad
d_{k\sigma }^{\dagger }=-\alpha _{k\sigma }^{\dagger }\sin \varphi
_{k}^{{}}+\beta _{k\sigma }^{\dagger }\cos \varphi _{k}^{{}}
\label{eq.hybrid}
\end{equation}

In the following, any operator $O$\ expressed through the hybridized
operators will be denoted by a tilde $\widetilde{O.}$ By choosing 
\begin{equation}
\tan \left( 2\varphi _{k}^{{}}\right) =-\frac{2zt_{k}^{cd}}{\varepsilon
_{k}^{c}-\varepsilon _{k}^{d}+z\left( t_{k}^{cc}-t_{k}^{dd}\right) }
\label{lin.7}
\end{equation}

$H_{el}^{{}}$ is brought to diagonal form $H_{el}^{{}}\Longrightarrow 
\widetilde{H}_{el}^{{}}=\sum_{k\sigma }\left( E_{k}^{\alpha }n_{k\sigma
}^{\alpha }+E_{k}^{\beta }n_{k\sigma }^{\beta }\right) $ , with the particle
energies in the hybridized bands given by: 
\begin{equation}
E_{k}^{\alpha }=\frac{1}{2}\left[ \varepsilon _{{}}^{c}+\varepsilon
_{{}}^{d}+z\left( t_{k}^{cc}+t_{k}^{dd}\right) \right] +\frac{1}{2}\sqrt{%
\left[ \varepsilon _{{}}^{c}-\varepsilon _{{}}^{d}+z\left(
t_{k}^{cc}-t_{k}^{dd}\right) \right] ^{2}+\left( 2zt_{k}^{cd}\right) ^{2}}
\label{lin.8}
\end{equation}
\begin{equation}
E_{k}^{\beta }=\frac{1}{2}\left[ \varepsilon _{{}}^{c}+\varepsilon
_{{}}^{d}+z\left( t_{k}^{cc}+t_{k}^{dd}\right) \right] -\frac{1}{2}\sqrt{%
\left[ \varepsilon _{{}}^{c}-\varepsilon _{{}}^{d}+z\left(
t_{k}^{cc}-t_{k}^{dd}\right) \right] ^{2}+\left( 2zt_{k}^{cd}\right) ^{2}}
\label{eq.hybE}
\end{equation}

The $\alpha $ and $\beta $ \ bands, at this stage completely decoupled by
the transformation of Eq.\ref{eq.hybrid}, represent in our model the $\sigma 
$ and $\pi $ bands of $MgB_{2}.$

\section{ The phononic Hamiltonian with anharmonic terms.}

Following \cite{Yildirim,choi} we assume that the purely phononic
Hamiltonian $H_{ph}^{{}}$ for $MgB_{2}$ has to include, apart from the usual
harmonic term, also a non-negligible quartic contribution. The anharmonicity
of the $MgB_{2}$ phonon modes has been analyzed in \cite{kunc}, showing that
two $E_{2g}$ modes, degenerate at the $\Gamma $ point, have anharmonicities
differing in the presence (mode labeled $a$ in \cite{kunc}), or absence
(mode $b$) , of a third-order term. In Ref.\cite{liu} a cubic term was
included in the development of the deformation energy. Its amplitude was
found to be of the same order as the linear term, i.e. about five times
smaller than the quartic term amplitude. It gives rise to terms non
conserving the phonon numbers, analogously to the linear SSH term. We'll
consider only the mode labeled $E_{2g}(b)$ in Ref.\cite{kunc}, which has no
third-order anharmonicity, both for short, and because the effect of the
dropped terms is similar, in amplitude and in type, to the one due to the
linear SSH\ term, which we keep. A similar assumption of neglecting the $%
E_{2g}(a)$ mode has been explicitly\cite{boeri} or implicitly\cite
{Yildirim,choi} made in other studies.

Under such assumptions, $H_{ph}^{{}}$ takes the form\cite{jones}: 
\begin{equation}
H_{ph}^{{}}=\sum_{q}\frac{P_{q}^{{}}P_{-q}^{{}}}{2M}+\frac{M}{2}%
\sum_{q}\Omega _{q}^{2}u_{q}^{{}}u_{-q}^{{}}+\frac{M^{2}}{4N}%
\sum_{qp}x_{qp}^{{}}\Omega _{q}^{2}\Omega
_{p}^{2}u_{q}^{{}}u_{-q}^{{}}u_{p}^{{}}u_{-p}^{{}}
\end{equation}

where $M$ \ is the Boron mass and $\Omega _{q}^{{}}$ is the frequency, at
the wavevector $q$ along the $\Gamma -A$ line, of the optical mode of $%
E_{2g} $ symmetry. The parameter $x_{qp}^{{}}$ expresses the strength of the
quartic term involving the wavevectors $\pm q$ and $\pm p.$ In $MgB_{2}$,
from Ref.\cite{Yildirim}, one can estimate $x_{qp}^{{}}\approx 7.8$ eV$%
^{-1}. $

By quantizing the phonon field according to the usual relations: 
\begin{equation}
u_{q}^{{}}=\sqrt{\frac{\hbar }{2M\Omega _{q}^{{}}}}\left( b_{-q}^{\dagger
}+b_{q}^{{}}\right) \qquad P_{q}^{{}}=i\sqrt{\frac{\hbar \Omega _{q}^{{}}}{2M%
}}\left( b_{-q}^{\dagger }-b_{q}^{{}}\right) \qquad L_{q}^{{}}=\sqrt{\frac{%
\hbar }{2M\Omega _{q}^{{}}}}  \label{ph.quant}
\end{equation}

the harmonic part becomes $\sum_{q}\hbar \Omega _{q}^{{}}\left(
b_{q}^{\dagger }b_{q}^{{}}+\frac{1}{2}\right) $. When quantizing the quartic
term, we neglect the terms with different numbers of creation and
destruction operators and keep the remaining four-operator products only
when diagonal. Namely, we approximate $b_{-q}^{\dagger }b_{q}^{\dagger
}b_{-p}^{{}}b_{p}^{{}}\approx \left( \delta _{p,q}^{{}}+\delta
_{p,-q}^{{}}\right) \nu _{q}^{{}}\nu _{-q}^{{}}$ , where $b_{q}^{\dagger
}b_{q}^{{}}=\nu _{q}^{{}}$. The quartic contribution then reduces to: 
\[
\sum_{qp}x_{qp}^{{}}\left( \frac{\hbar \Omega _{q}^{{}}}{4}\right) \left( 
\frac{\hbar \Omega _{p}^{{}}}{4}\right) \left( b_{-q}^{\dagger
}+b_{q}^{{}}\right) \left( b_{q}^{\dagger }+b_{-q}^{{}}\right) \left(
b_{-p}^{\dagger }+b_{p}^{{}}\right) \left( b_{p}^{\dagger
}+b_{-p}^{{}}\right) \approx 
\]

\[
\approx 4\sum_{q}\left( \frac{\hbar \Omega _{q}^{{}}}{4}\right) \left( \frac{%
1}{2}+\nu _{q}^{{}}\right) \sum_{p}x_{qp}^{{}}\left( \frac{\hbar \Omega
_{p}^{{}}}{4}\right) \left( 1+\delta _{qp}^{{}}\right) 
\]
\[
+4\sum_{qp}x_{qp}^{{}}\left( \frac{\hbar \Omega _{p}^{{}}}{4}\right) \left( 
\frac{\hbar \Omega _{q}^{{}}}{4}\right) \nu _{q}^{{}}\nu _{p}^{{}}\left(
1+\delta _{q,-p}^{{}}\right) +2\sum_{q}\left( \frac{\hbar \Omega _{q}^{{}}}{4%
}\right) \left( b_{-q}^{\dagger }b_{q}^{\dagger
}+b_{-q}^{{}}b_{q}^{{}}\right) \sum_{p}x_{qp}^{{}}\left( \frac{\hbar \Omega
_{p}^{{}}}{4}\right) 
\]
\begin{equation}
-2\sum_{qp}x_{qp}^{{}}\left( \frac{\hbar \Omega _{q}^{{}}}{4}\right) \left( 
\frac{\hbar \Omega _{p}^{{}}}{4}\right) +\sum_{qp}x_{qp}^{{}}\left( \frac{%
\hbar \Omega _{q}^{{}}}{4}\right) \left( \frac{\hbar \Omega _{p}^{{}}}{4}%
\right)  \label{5.2}
\end{equation}

The product $\nu _{q}^{{}}\nu _{p}^{{}}$\ is approximated in the MFA
fashion, i.e. $\nu _{q}^{{}}\nu _{p}^{{}}\approx \nu _{q}^{{}}\langle \nu
_{p}^{{}}\rangle +\langle \nu _{q}^{{}}\rangle \nu _{p}^{{}}-\langle \nu
_{p}^{{}}\rangle \langle \nu _{q}^{{}}\rangle $ . Putting together the
constant terms, we can rewrite Eq.\ref{5.2} as: 
\[
\sum_{q}\hbar \Omega _{q}^{{}}\left( \frac{1}{2}+\nu _{q}^{{}}\right) \left[ 
\frac{1}{N}\sum_{p}x_{qp}^{{}}\left( \frac{\hbar \Omega _{p}^{{}}}{2}\right)
\left( \frac{1}{2}+\langle \nu _{p}^{{}}\rangle \right) \left( 1+\delta
_{q,-p}^{{}}\right) \right] 
\]
\begin{equation}
+\sum_{q}\hbar \Omega _{q}^{{}}\left( b_{-q}^{\dagger }b_{q}^{\dagger
}+b_{-q}^{{}}b_{q}^{{}}\right) \left[ \frac{1}{N}\sum_{p}x_{qp}^{{}}\left( 
\frac{\hbar \Omega _{p}^{{}}}{8}\right) \right] +{\rm const.}  \label{8.1}
\end{equation}

Adding the harmonic contribution and defining 
\begin{equation}
X_{q}^{{}}\equiv 1+\frac{1}{N}\sum_{p}x_{qp}^{{}}\left( \frac{\hbar \Omega
_{p}^{{}}}{2}\right) \left( \frac{1}{2}+\langle \nu _{p}^{{}}\rangle \right)
\left( 1+\delta _{q,-p}^{{}}\right)  \label{eqXq}
\end{equation}

we obtain the purely phononic Hamiltonian as: 
\begin{equation}
H_{ph}=\sum_{q}\hbar \Omega _{q}^{{}}X_{q}^{{}}\left( \frac{1}{2}+\nu
_{q}^{{}}\right) +\sum_{q}\hbar \Omega _{q}^{{}}\left( b_{-q}^{\dagger
}b_{q}^{\dagger }+b_{-q}^{{}}b_{q}^{{}}\right) \left[ \frac{1}{N}%
\sum_{p}x_{qp}^{{}}\left( \frac{\hbar \Omega _{p}^{{}}}{8}\right) \right] +%
{\rm const.}  \label{8.2b}
\end{equation}

This form can be diagonalized by a \ ''squeezing '' transformation\cite
{squeeze} $e^{S}\equiv \exp \left[ -\sum_{q}\eta _{q}^{{}}\left(
b_{-q}^{\dagger }b_{q}^{\dagger }-b_{-q}^{{}}b_{q}^{{}}\right) \right] $
under the condition that 
\begin{equation}
\tanh \left( 2\eta _{q}^{{}}\right) =-\frac{1}{X_{q}^{{}}}\left( \frac{1}{N}%
\right) \sum_{p}x_{qp}^{{}}\left( \frac{\hbar \Omega _{p}^{{}}}{4}\right)
\label{10.4}
\end{equation}

Notice that Eq.\ref{10.4} yields $\eta _{q}^{{}}<0.$ The diagonalized
Hamiltonian $e^{S}H_{ph}e^{-S}$ can now be written as: 
\begin{equation}
e^{S}H_{ph}e^{-S}=\sum_{q}\hbar \Omega _{q}^{{}}\left[ X_{q}^{{}}\cosh
\left( 2\eta _{q}^{{}}\right) +2\sinh \left( 2\eta _{q}^{{}}\right) \left( 
\frac{1}{N}\right) \sum_{p}x_{qp}^{{}}\frac{\hbar \Omega _{p}^{{}}}{8}\right]
\left( b_{q}^{\dagger }b_{q}^{{}}+\frac{1}{2}\right) +{\rm const.}
\label{11.3}
\end{equation}

By substituting $\eta _{q}^{{}}$ from Eq.\ref{10.4} into Eq.\ref{11.3}, the
renormalized frequency $\omega _{q}^{{}}$ of the harmonic Hamiltonian for
the squeezed phonons is written explicitly as: 
\begin{equation}
\omega _{q}^{{}}=\Omega _{q}^{{}}X_{q}^{{}}\left[ \sqrt{1-\tanh ^{2}\left(
2\eta _{q}^{{}}\right) }\right]  \label{SQomega}
\end{equation}

where $\Omega _{q}^{{}}X_{q}^{{}}$ is the phonon frequency entering the
quadratic part of the unsqueezed phononic Hamiltonian (see. Eq.\ref{8.2b}).
It is not the true bare frequency, because, from Eq.\ref{eqXq}, $%
X_{q}^{{}}-1 $ yields the contribution from the diagonal part of the quartic
terms treated in MFA, so that $\Omega _{q}^{{}}X_{q}^{{}}$ already contains
some effects of anharmonicity, analogous to those taken into account, e.g.
in Ref. \cite{Yildirim}. Thus, Eq.\ref{SQomega} shows that $\omega _{q}^{{}}$
is increased (hardened) with respect to the ''harmonic frequency'' $\Omega
_{q}^{{}}$ by $X_{q}^{{}}>1.$, but the squeezing effect, taking account of
the two-phonon terms terms previously\cite{Yildirim} neglected, counteracts
the hardening. According to Refs.\cite{liu,Yildirim,choi} however, the
squeezing effect is not strong enough for an overall softening to result.

\section{The linear electron-phonon interaction.}

The linear part of the SSH\ electron-phonon interaction is written, in the
real space representation symmetrized with respect to the site indexes, as 
\[
H_{ep}^{(1)}=\frac{1}{2}\sum_{l\langle j\rangle \sigma }\left[
g_{lj}^{cc}\left( c_{l\sigma }^{\dagger }c_{j\sigma }^{{}}+c_{j\sigma
}^{\dagger }c_{l\sigma }^{{}}\right) +g_{lj}^{dd}\left( d_{l\sigma
}^{\dagger }d_{j\sigma }^{{}}+d_{j\sigma }^{\dagger }d_{l\sigma
}^{{}}\right) \right] \left( u_{l}^{{}}-u_{j}^{{}}\right) 
\]
\begin{equation}
+\frac{1}{2}\sum_{l\langle j\rangle \sigma }g_{lj}^{cd}\left( d_{l\sigma
}^{\dagger }c_{j\sigma }^{{}}+c_{j\sigma }^{\dagger }d_{l\sigma
}^{{}}+d_{j\sigma }^{\dagger }c_{l\sigma }^{{}}+c_{l\sigma }^{\dagger
}d_{j\sigma }^{{}}\right) \left( u_{l}^{{}}-u_{j}^{{}}\right)  \label{lin.1}
\end{equation}

where $g_{lj}^{\mu \nu }=\partial t_{lj}^{\mu \nu }/\partial \left(
u_{l}^{{}}-u_{j}^{{}}\right) |_{0}=-g_{jl}^{\mu \nu }$ with $\mu ,\nu =c,d$,
are the coupling constants, and the $1/2$ factor avoids double counting.

The Fourier-transformed form of Eq.\ref{lin.1} is written in terms of $%
g_{k}^{\mu \nu }=(1/z)\sum_{\langle j\rangle }g_{lj}^{\mu \nu }\exp (ik\cdot
\Delta _{lj}^{{}})$ (where $\Delta _{lj}^{{}}$ is the vector connecting the
sites $l$ and $j)$ which we combine in the definition of the coupling
strength $\gamma _{k,k-q}^{\mu \nu }$ according to: 
\begin{equation}
\frac{z}{2}\left( g_{k-q}^{\mu \nu }+g_{k}^{\mu \nu }-g_{-(k-q)}^{\mu \nu
}-g_{-k}^{\mu \nu }\right) =i\sum_{\langle j\rangle }g_{lj}^{\mu \nu
}\left\{ \sin \left[ \left( k-q\right) \cdot \Delta _{lj}^{{}}\right] -\sin %
\left[ k\cdot \Delta _{lj}^{{}}\right] \right\} \equiv \gamma _{k,k-q}^{\mu
\nu }/L_{q}^{{}}  \label{lin.3}
\end{equation}

Eq.\ref{lin.3} is the simplest possible form of the bond-stretching
electron-phonon interaction which includes the relevant physics. It is
adequate for a qualitative discussion, but it is not good enough for a
quantitative study.

Quantization of the phonons according to Eq.\ref{ph.quant} leads to: 
\[
H_{ep}^{(1)}= 
\]
\begin{equation}
=\frac{1}{\sqrt{N}}\sum_{kq\sigma }\left[ \gamma _{k,k-q}^{cc}c_{k\sigma
}^{\dagger }c_{k-q\sigma }^{{}}+\gamma _{k,k-q}^{dd}d_{k\sigma }^{\dagger
}d_{k-q\sigma }^{{}}+\gamma _{k,k-q}^{cd}\left( c_{k\sigma }^{\dagger
}d_{k-q\sigma }^{{}}+d_{k\sigma }^{\dagger }c_{k-q\sigma }^{{}}\right) %
\right] \left( b_{-q}^{\dagger }+b_{q}^{{}}\right)  \label{eq.SSH1d}
\end{equation}

When transformed to the hybridized fermion representation $H_{ep}^{(1)}$
reads: 
\[
\widetilde{H}_{ep}^{(1)}= 
\]
\begin{equation}
=\frac{1}{\sqrt{N}}\sum_{kq\sigma }\left[ \Gamma _{k,k-q}^{\alpha \alpha
}\alpha _{k\sigma }^{\dagger }\alpha _{k-q,\sigma }^{{}}+\Gamma
_{k,k-q}^{\beta \beta }\beta _{k\sigma }^{\dagger }\beta _{k-q,\sigma
}^{{}}+\Gamma _{k,k-q}^{\alpha \beta }\alpha _{k\sigma }^{\dagger }\beta
_{k-q,\sigma }^{{}}+\Gamma _{k,k-q}^{\beta \alpha }\beta _{k\sigma
}^{\dagger }\alpha _{k-q,\sigma }^{{}}\right] \left( b_{-q}^{\dagger
}+b_{q}^{{}}\right)  \label{lin.9}
\end{equation}

where the effective couplings are defined as: 
\begin{equation}
\Gamma _{k,k-q}^{\alpha \alpha }=\gamma _{k,k-q}^{cc}\cos \varphi
_{k}^{{}}\cos \varphi _{k-q}^{{}}+\gamma _{k,k-q}^{dd}\sin \varphi
_{k}^{{}}\sin \varphi _{k-q}^{{}}-\gamma _{k,k-q}^{cd}\sin \left( \varphi
_{k}^{{}}+\varphi _{k-q}^{{}}\right)
\end{equation}

\begin{equation}
\Gamma _{k,k-q}^{\beta \beta }=\gamma _{k,k-q}^{cc}\sin \varphi
_{k}^{{}}\sin \varphi _{k-q}^{{}}+\gamma _{k,k-q}^{dd}\cos \varphi
_{k}^{{}}\cos \varphi _{k-q}^{{}}+\gamma _{k,k-q}^{cd}\sin \left( \varphi
_{k}^{{}}+\varphi _{k-q}^{{}}\right)
\end{equation}

\begin{equation}
\Gamma _{k,k-q}^{\alpha \beta }=\gamma _{k,k-q}^{cc}\cos \varphi
_{k}^{{}}\sin \varphi _{k-q}^{{}}-\gamma _{k,k-q}^{dd}\sin \varphi
_{k}^{{}}\cos \varphi _{k-q}^{{}}+\gamma _{k,k-q}^{cd}\cos \left( \varphi
_{k}^{{}}+\varphi _{k-q}^{{}}\right)
\end{equation}

\begin{equation}
\Gamma _{k,k-q}^{\beta \alpha }=\gamma _{k,k-q}^{cc}\sin \varphi
_{k}^{{}}\cos \varphi _{k-q}^{{}}-\gamma _{k,k-q}^{dd}\cos \varphi
_{k}^{{}}\sin \varphi _{k-q}^{{}}+\gamma _{k,k-q}^{cd}\cos \left( \varphi
_{k}^{{}}+\varphi _{k-q}^{{}}\right)
\end{equation}

\section{The quadratic electron-phonon interaction.}

According to Refs.\cite{liu,Yildirim,choi}, the electron-phonon Hamiltonian
has to include also a quadratic term, which we write in real space in
symmetrized form, as: 
\[
H_{ep}^{(2)}=\frac{1}{2}\sum_{l\langle j\rangle \sigma }\left[
f_{lj}^{cc}\left( c_{l\sigma }^{\dagger }c_{j\sigma }^{{}}+c_{j\sigma
}^{\dagger }c_{l\sigma }^{{}}\right) +f_{lj}^{dd}\left( d_{l\sigma
}^{\dagger }d_{j\sigma }^{{}}+d_{j\sigma }^{\dagger }d_{l\sigma
}^{{}}\right) \right] \left( u_{l}^{{}}-u_{j}^{{}}\right) ^{2} 
\]
\begin{equation}
+\frac{1}{2}\sum_{l\langle j\rangle \sigma }f_{lj}^{cd}\left( d_{l\sigma
}^{\dagger }c_{j\sigma }^{{}}+c_{j\sigma }^{\dagger }d_{l\sigma
}^{{}}+d_{j\sigma }^{\dagger }c_{l\sigma }^{{}}+c_{l\sigma }^{\dagger
}d_{j\sigma }^{{}}\right) \left( u_{l}^{{}}-u_{j}^{{}}\right) ^{2}
\label{eq9.0}
\end{equation}

where $f_{lj}^{\mu \nu }=\partial ^{2}t_{lj}^{\mu \nu }/\partial \left(
u_{l}^{{}}-u_{j}^{{}}\right) ^{2}|_{0}$ $=f_{jl}^{\mu \nu }$, with $\mu ,\nu
=c,d$. We also develop $\left( u_{l}^{{}}-u_{j}^{{}}\right)
^{2}=u_{l}^{2}+u_{j}^{2}-u_{l}^{{}}u_{j}^{{}}-u_{j}^{{}}u_{l}^{{}}.$ By
defining $f_{k}^{\mu \nu }\equiv z^{-1}\sum_{\langle j\rangle }f_{lj}^{\mu
\nu }e^{ik\Delta _{lj}^{{}}}$ and introducing the coefficients $F_{kpq}^{\mu
\nu }\equiv \left( z/2\right) \left( f_{p}^{\mu \nu }+f_{k}^{\mu \nu
}-f_{k-q}^{\mu \nu }-f_{p+q}^{\mu \nu }\right) $ $,$ the Fourier transform
reads:

\[
H_{ep}^{(2)}=\frac{1}{N}\sum_{kpq\sigma }\left[ F_{kpq}^{cc}\left(
c_{k\sigma }^{\dagger }c_{p\sigma }^{{}}+c_{-p\sigma }^{\dagger }c_{-k\sigma
}^{{}}\right) +F_{kpq}^{dd}\left( d_{k\sigma }^{\dagger }d_{p\sigma
}^{{}}+d_{-p\sigma }^{\dagger }d_{-k\sigma }^{{}}\right) \right]
u_{q}^{{}}u_{k-p-q}^{{}}+ 
\]
\begin{equation}
+\frac{1}{N}\sum_{kpq\sigma }F_{kpq}^{cd}\left( d_{k\sigma }^{\dagger
}c_{p\sigma }^{{}}+c_{k\sigma }^{\dagger }d_{p\sigma }^{{}}+d_{-p\sigma
}^{\dagger }c_{-k\sigma }^{{}}+c_{-p\sigma }^{\dagger }d_{-k\sigma
}^{{}}\right) u_{q}^{{}}u_{k-p-q}^{{}}  \label{eq.SSH2a}
\end{equation}

When quantizing the deformations according to Eq.\ref{ph.quant}, we shall
take into account only the diagonal terms and those which can be
diagonalized by squeezing, by enforcing $k=p.$ Then $%
u_{q}^{{}}u_{k-p-q}^{{}} $ reduces to: 
\begin{equation}
u_{q}^{{}}u_{k-p-q}^{{}}\Longrightarrow u_{q}^{{}}u_{-q}^{{}}\delta
_{pk}^{{}}=\delta _{pk}^{{}}L_{q}^{2}\left( b_{-q}^{\dagger }b_{q}^{\dagger
}+b_{q}^{{}}b_{-q}^{{}}+\nu _{q}^{{}}+\nu _{-q}^{{}}+1\right)  \label{20.3}
\end{equation}

Let us stress that our aim is to show that there are some contributions to $%
H_{ep}^{(2)}$ which provide an effective inter-band coupling. We do not
claim to be able to treat all the terms in $H_{ep}^{(2)}$: we just want to
select the subset of ''hot\ ''terms. Under this approximation, by using $%
F_{k,q}^{\mu \nu }=F_{-k,-q}^{\mu \nu }$ , Eq.\ref{eq.SSH2a} reduces to:

\[
H_{ep}^{(2)}\approx 2\frac{1}{N}\sum_{kq\sigma }L_{q}^{2}\left[
F_{k,q}^{cc}c_{k\sigma }^{\dagger }c_{k\sigma }^{{}}+F_{kq}^{dd}d_{k\sigma
}^{\dagger }d_{k\sigma }^{{}}+F_{kq}^{cd}\left( d_{k\sigma }^{\dagger
}c_{p\sigma }^{{}}+c_{k\sigma }^{\dagger }d_{p\sigma }^{{}}\right) \right]
\otimes 
\]
\begin{equation}
\otimes \left( b_{-q}^{\dagger }b_{q}^{\dagger }+b_{q}^{{}}b_{-q}^{{}}+\nu
_{q}^{{}}+\nu _{-q}^{{}}+1\right)  \label{eq.SSH2b}
\end{equation}

Let' s now pass to the hybridized band picture, through the transformation
of Eq.\ref{eq.hybrid}. If we define for short the energies: 
\begin{equation}
F_{kq}^{\alpha \alpha }=2L_{q}^{2}\left[ F_{kq}^{cc}\cos ^{2}\varphi
_{k}^{{}}+F_{kq}^{dd}\sin ^{2}\varphi _{k}^{{}}-F_{kq}^{cd}\sin \left(
2\varphi _{k}^{{}}\right) \right]  \label{18.1a}
\end{equation}

\begin{equation}
F_{kq}^{\beta \beta }=2L_{q}^{2}\left[ F_{kq}^{cc}\sin ^{2}\varphi
_{k}^{{}}+F_{kq}^{dd}\cos ^{2}\varphi _{k}^{{}}+F_{kq}^{cd}\sin \left(
2\varphi _{k}^{{}}\right) \right]  \label{18.1b}
\end{equation}

\begin{equation}
F_{kq}^{\alpha \beta }=L_{q}^{2}\left[ \left( F_{kq}^{cc}-F_{kq}^{dd}\right)
\sin \left( 2\varphi _{k}^{{}}\right) +2F_{kq}^{cd}\cos \left( 2\varphi
_{k}^{{}}\right) \right]
\end{equation}

where 
\begin{equation}
F_{kq}^{\mu \nu }=\sum_{\langle j\rangle }f_{lj}^{\mu \nu }\left\{ \cos
\left( k\Delta _{lj}^{{}}\right) \left[ 1-\cos \left( q\Delta
_{lj}^{{}}\right) \right] \right\} \qquad \left( \mu ,\nu =c,d\right)
\end{equation}

then Eq.\ref{eq.SSH2b} can be rewritten compactly as: 
\[
\widetilde{H}_{ep}^{(2)}=\frac{1}{N}\sum_{kq\sigma }\left[ n_{k\sigma
}^{\alpha }F_{kq}^{\alpha \alpha }+n_{k\sigma }^{\beta }F_{kq}^{\beta \beta
}+\left( \alpha _{k\sigma }^{\dagger }\beta _{k\sigma }^{{}}+\beta _{k\sigma
}^{\dagger }\alpha _{p\sigma }^{{}}\right) F_{kq}^{\alpha \beta }\right]
\otimes 
\]
\begin{equation}
\otimes \left( b_{-q}^{\dagger }b_{q}^{\dagger }+b_{q}^{{}}b_{-q}^{{}}+\nu
_{q}^{{}}+\nu _{-q}^{{}}+1\right)
\end{equation}

\section{\protect\bigskip\ The electron-phonon Hamiltonian in the squeezed
phonon representation.}

Let us now introduce the squeezed phonon representation also for $\widetilde{%
H}_{ep}^{(1)}+\widetilde{H}_{ep}^{(2)}$. By using the relation $e^{S}\left(
b_{-q}^{\dagger }+b_{q}^{{}}\right) e^{-S}=e^{\eta _{q}^{{}}}\left(
b_{-q}^{\dagger }+b_{q}^{{}}\right) $ the linear coupling term becomes: 
\[
e^{S}\widetilde{H}_{ep}^{(1)}e^{-S}= 
\]
\begin{equation}
=\frac{1}{\sqrt{N}}\sum_{kq\sigma }e^{\eta _{q}^{{}}}\left[ \Gamma
_{k,k-q}^{\alpha \alpha }\alpha _{k\sigma }^{\dagger }\alpha _{k-q,\sigma
}^{{}}+\Gamma _{k,k-q}^{\beta \beta }\beta _{k\sigma }^{\dagger }\beta
_{k-q,\sigma }^{{}}+\Gamma _{k,k-q}^{\alpha \beta }\alpha _{k\sigma
}^{\dagger }\beta _{k-q,\sigma }^{{}}+\Gamma _{k,k-q}^{\beta \alpha }\beta
_{k\sigma }^{\dagger }\alpha _{k-q,\sigma }^{{}}\right] \left(
b_{-q}^{\dagger }+b_{q}^{{}}\right)
\end{equation}

Then the linear coupling has a reduce amplitude, as $\eta _{q}^{{}}<0$ (see
Eq.\ref{10.4}), consistently with the numerical analysis of Ref. \cite{choi}.

For the quadratic part $\widetilde{H}_{ep}^{(2)}$ we get:

\[
e^{S}\widetilde{H}_{ep}^{(2)}e^{-S}=\frac{1}{N}\sum_{kq}\left[
F_{kq}^{\alpha \alpha }n_{k\sigma }^{\alpha }+F_{kq}^{\beta \beta
}n_{k\sigma }^{\beta }+F_{kq}^{\alpha \beta }\left( \alpha _{k\sigma
}^{\dagger }\beta _{k\sigma }^{{}}+\beta _{k\sigma }^{\dagger }\alpha
_{k\sigma }^{{}}\right) \right] \otimes 
\]
\begin{equation}
\otimes e^{2\eta _{q}^{{}}}\left( b_{-q}^{\dagger }b_{q}^{\dagger
}+b_{q}^{{}}b_{-q}^{{}}+\nu _{q}^{{}}+\nu _{-q}^{{}}+1\right)  \label{23.1}
\end{equation}

In Eq.\ref{23.1} we can decouple the electronic and phononic terms in MFA.
Indeed, assuming 
\begin{equation}
\langle b_{-q}^{\dagger }b_{q}^{\dagger }+b_{q}^{{}}b_{-q}^{{}}+\nu
_{q}^{{}}+\nu _{-q}^{{}}+1\rangle \approx 2\langle \nu _{q}^{{}}\rangle +1
\end{equation}

and defining 
\begin{equation}
\Phi _{q}^{{}}=\frac{1}{N}\sum_{k\sigma }\left[ F_{kq}^{\alpha \alpha
}\langle n_{k\sigma }^{\alpha }\rangle +2F_{kq}^{\beta \beta }\langle
n_{k\sigma }^{\beta }\rangle +F_{kq}^{\alpha \beta }\left( \langle \alpha
_{k\sigma }^{\dagger }\beta _{k\sigma }^{{}}\rangle +\langle \beta _{k\sigma
}^{\dagger }\alpha _{k\sigma }^{{}}\rangle \right) \right] =\Phi _{-q}^{{}}
\label{eq.Phiq}
\end{equation}

yields:

\[
e^{S}\widetilde{H}_{ep}^{\left( 2\right) }e^{-S}\approx \frac{1}{N}%
\sum_{kq\sigma }\left[ F_{kq}^{\alpha \alpha }n_{k\sigma }^{\alpha
}+F_{kq}^{\beta \beta }n_{k\sigma }^{\beta }+F_{kq}^{\alpha \beta }\left(
\alpha _{k\sigma }^{\dagger }\beta _{k\sigma }^{{}}+\beta _{k\sigma
}^{\dagger }\alpha _{k\sigma }^{{}}\right) \right] e^{2\eta _{q}^{{}}}\left(
2\langle \nu _{q}^{{}}\rangle +1\right) 
\]
\begin{equation}
+\sum_{q}e^{2\eta _{q}^{{}}}\Phi _{q}^{{}}\left( b_{-q}^{\dagger
}b_{q}^{\dagger }+b_{q}^{{}}b_{-q}^{{}}+\nu _{q}^{{}}+\nu
_{-q}^{{}}+1\right) +{\rm const.}  \label{R1.5}
\end{equation}

\bigskip By writing $\sum_{q}\hbar \omega _{q}^{{}}\left( \nu _{q}^{{}}+%
\frac{1}{2}\right) =\frac{1}{2}\sum_{q}\hbar \omega _{q}^{{}}\left( \nu
_{q}^{{}}+\nu _{-q}^{{}}+1\right) $ \ and reordering the Hamiltonian in the
hybridized basis we get: 
\begin{equation}
e^{S}\widetilde{H}e^{-S}=\sum_{k\sigma }E_{k}^{\alpha }n_{k\sigma }^{\alpha
}+\sum_{k\sigma }E_{k}^{\beta }n_{k\sigma }^{\beta }+\sum_{q}\hbar \omega
_{q}^{{}}\left( \nu _{q}^{{}}+\frac{1}{2}\right) +e^{S}\widetilde{H}%
_{ep}^{\left( 1\right) }e^{-S}+e^{S}\widetilde{H}_{ep}^{\left( 2\right)
}e^{-S}  \label{R1.1}
\end{equation}

Inserting $e^{S}\widetilde{H}_{ep}^{\left( 2\right) }e^{-S}$ from \ Eq.\ref
{R1.5} into Eq.\ref{R1.1} and reordering yields: 
\begin{equation}
e^{S}\widetilde{H}e^{-S}={\cal H}_{el.}^{diag}{\cal +H}_{el}^{hyb}{\cal +H}%
_{ph}^{{}}+e^{S}H_{ep}^{\left( 1\right) }e^{-S}+{\rm const.}
\label{eq.Hamtilde}
\end{equation}

where the diagonal electronic Hamiltonian is: 
\begin{equation}
{\cal H}_{el.}^{diag}\equiv \sum_{k\sigma }\left[ E_{k}^{\alpha }+\frac{1}{N}%
\sum_{q}F_{kq}^{\alpha \alpha }e^{2\eta _{q}^{{}}}\left( 2\langle \nu
_{q}^{{}}\rangle +1\right) \right] n_{k\sigma }^{\alpha }+\sum_{k\sigma }%
\left[ E_{k}^{\beta }+\frac{1}{N}\sum_{q}F_{kq}^{\beta \beta }e^{2\eta
_{q}^{{}}}\left( 2\langle \nu _{q}^{{}}\rangle +1\right) \right] n_{k\sigma
}^{\beta }  \label{eq.Heldiag}
\end{equation}

and describes a phonon-depending renormalization of the band energies. The
hybrid electronic Hamiltonian is: 
\begin{equation}
{\cal H}_{el}^{hyb}\equiv \frac{1}{N}\sum_{kq\sigma }e^{2\eta
_{q}^{{}}}F_{kq}^{\alpha \beta }\left( 2\langle \nu _{q}^{{}}\rangle
+1\right) \left( \alpha _{k\sigma }^{\dagger }\beta _{k\sigma }^{{}}+\beta
_{k\sigma }^{\dagger }\alpha _{k\sigma }^{{}}\right)  \label{eq.Hamelhyb}
\end{equation}

It \ represents a phonon-depending band hybridization term. This is, we
believe, the term responsible for the coupling between the bands in $MgB_{2}$
which results in a single critical temperature for both gaps. We would like
to point out that, in the limit of small inter-bare-band hopping, i.e. $%
t_{lj}^{cd}/t_{lj}^{cc}\rightarrow 0,$ and at zero temperature, where $%
\langle \nu _{q}^{{}}\rangle $ can be neglected, one finds: 
\begin{equation}
\lim_{t_{lj}^{cd}/t_{lj}^{cc}\rightarrow 0}e^{2\eta _{q}^{{}}}F_{kq}^{\alpha
\beta }=2L_{q}^{2}e^{2\eta _{q}^{{}}}F_{kq}^{cd}=2L_{q}^{2}e^{2\eta
_{q}^{{}}}\sum_{\langle j\rangle }f_{lj}^{cd}\left\{ \cos \left( k\Delta
_{lj}^{{}}\right) \left[ 1-\cos \left( q\Delta _{lj}^{{}}\right) \right]
\right\}
\end{equation}

Apart from geometric factors, this amplitude depends only on the intensity
of the squeezing (through $e^{2\eta _{q}^{{}}})$ and on the strength of the
quadratic inter-band SSH electron-phonon coupling $f_{lj}^{cd}$. As $%
f_{lj}^{cd}$ is a second derivative of $t_{lj}^{cd},$ it can be
non-negligible even if $t_{lj}^{cd}$ itself is very small. Different
evaluations of $f_{lj}^{cd}$\cite{mazin} \ all agree that in $MgB_{2}$ it
has an appreciable value. More specifically, an effective two-band model
derived from first-principle calculations\cite{liu}\ yields $\lambda
_{\sigma \pi }/$\ $\lambda _{\sigma \sigma }=0.21$ and{\it \ }$\lambda _{\pi
\sigma }/\lambda _{\sigma \sigma }=0.15${\it \ }\cite{golubov}.

At nonzero temperatures, the hybridization amplitude gets an additional
contribution proportional to $2\langle \nu _{q}^{{}}\rangle $, hence it
increases with temperature, consistently with the findings of Ref. \cite
{dahm}

The purely phononic term 
\begin{equation}
{\cal H}_{ph}^{{}}\equiv \sum_{q}\left( \frac{1}{2}\hbar \omega
_{q}^{{}}+e^{2\eta _{q}^{{}}}\Phi _{q}^{{}}\right) \left( \nu _{q}^{{}}+\nu
_{-q}^{{}}+1\right) +\sum_{q}e^{2\eta _{q}^{{}}}\Phi _{q}^{{}}\left(
b_{-q}^{\dagger }b_{q}^{\dagger }+b_{q}^{{}}b_{-q}^{{}}\right)  \label{R1.6}
\end{equation}

is diagonalized by a second squeezing transformation $e^{T}\equiv \exp \left[
-\sum_{q}\vartheta _{q}^{{}}\left( b_{-q}^{\dagger }b_{q}^{\dagger
}-b_{-q}^{{}}b_{q}^{{}}\right) \right] $ with the value of $\vartheta
_{q}^{{}}$ set by: 
\begin{equation}
\tanh \left( 2\vartheta _{q}^{{}}\right) =\frac{\hbar \omega _{q}^{{}}}{%
\hbar \omega _{q}^{{}}+2e^{2\eta _{q}^{{}}}\Phi _{q}^{{}}}-1
\label{eq.thetaq}
\end{equation}

Notice that the sign of $\vartheta _{q}^{{}}$ is opposite to the sign of \ $%
\Phi _{q}^{{}}.$ Also, as the relevant phonons are optical ones, $\hbar
\omega _{q}^{{}}$ never vanishes in the Brillouin zone, then $\tanh \left(
2\vartheta _{q}^{{}}\right) \neq -1$ , and $\vartheta _{q}^{{}}$ is always
well defined. Due to the presence of $\Phi _{q}^{{}}$ in Eq.\ref{eq.thetaq},
the parameter $\vartheta _{q}^{{}}$ depends on the band-filling factors $%
\langle n_{k\sigma }^{\alpha (\beta )}\rangle $ and on the band
hybridization $\langle \alpha _{k\sigma }^{\dagger }\beta _{k\sigma
}^{{}}+\beta _{k\sigma }^{\dagger }\alpha _{k\sigma }^{{}}\rangle $.

The diagonalized free-phonon Hamiltonian reads therefore: 
\begin{equation}
e^{T}{\cal H}_{ph}^{{}}e^{-T}=\sum_{q}\left[ \hbar \omega _{q}^{{}}\cosh
\left( 2\vartheta _{q}^{{}}\right) +2e^{2\eta _{q}^{{}}}e^{2\vartheta
_{q}^{{}}}\Phi _{q}^{{}}\right] \nu _{q}^{{}}+{\rm const.}=\sum_{q}\hbar
\varpi _{q}^{{}}\nu _{q}^{{}}+{\rm const.}  \label{R5.4}
\end{equation}

We have obtained a band-filling and hybridization-depending renormalization
of the phonon frequencies:

\begin{equation}
\hbar \varpi _{q}^{{}}=\hbar \omega _{q}^{{}}\cosh \left( 2\vartheta
_{q}^{{}}\right) +2e^{2\eta _{q}^{{}}}e^{2\vartheta _{q}^{{}}}\Phi _{q}^{{}}
\label{eq.46}
\end{equation}

The final Hamiltonian therefore reads:

\[
e^{T}{\cal H}e^{-T}=\sum_{k\sigma }\left( {\cal E}_{k}^{\alpha }n_{k\sigma
}^{\alpha }+{\cal E}_{k}^{\beta }n_{k\sigma }^{\beta }\right) +\frac{1}{N}%
\sum_{kq\sigma }\left[ e^{2\eta _{q}^{{}}}F_{kq}^{\alpha \beta }\left(
2\langle \nu _{q}^{{}}\rangle +1\right) \right] \left( \alpha _{k\sigma
}^{\dagger }\beta _{k\sigma }^{{}}+\beta _{k\sigma }^{\dagger }\alpha
_{k\sigma }^{{}}\right) 
\]
\begin{equation}
+\sum_{q}\hbar \varpi _{q}^{{}}\nu _{q}^{{}}+e^{T}e^{S}\widetilde{H}%
_{ep}^{(1)}e^{-S}e^{-T}+{\rm const.}  \label{eq.47}
\end{equation}

where 
\begin{equation}
{\cal E}_{k}^{\zeta }=E_{k}^{\zeta }+\frac{1}{N}\sum_{q}F_{kq}^{\zeta \zeta
}e^{2\eta _{q}^{{}}}\left( 2\langle \nu _{q}^{{}}\rangle +1\right) \qquad
\left( \zeta =\alpha ,\beta \right)
\end{equation}

This has the shape of a standard (i.e. harmonic and linear) SSH\ Hamiltonian
for two hybridizing bands. However, also the linear SSH term has acquired a
band-filling and hybridization dependence, because it now reads: 
\[
e^{T}e^{S}\widetilde{H}_{ep}^{(1)}e^{-S}e^{-T}= 
\]
\begin{equation}
=\frac{1}{\sqrt{N}}\sum_{kq\sigma }e^{\eta _{q}^{{}}+\vartheta _{q}}\left[
\Gamma _{k,k-q}^{\alpha \alpha }\alpha _{k\sigma }^{\dagger }\alpha
_{k-q,\sigma }^{{}}+\Gamma _{k,k-q}^{\beta \beta }\beta _{k\sigma }^{\dagger
}\beta _{k-q,\sigma }^{{}}+\Gamma _{k,k-q}^{\alpha \beta }\alpha _{k\sigma
}^{\dagger }\beta _{k-q,\sigma }^{{}}+\Gamma _{k,k-q}^{\beta \alpha }\beta
_{k\sigma }^{\dagger }\alpha _{k-q,\sigma }^{{}}\right] \left(
b_{-q}^{\dagger }+b_{q}^{{}}\right)  \label{eq.49}
\end{equation}

Therefore also the phonon linewidths, due to the SSH interaction, will
depend , through $\exp (\vartheta _{q}^{{}})$, on $\langle n_{k\sigma
}^{\alpha (\beta )}\rangle $ and $\langle \alpha _{k\sigma }^{\dagger }\beta
_{k\sigma }^{{}}+\beta _{k\sigma }^{\dagger }\alpha _{k\sigma }^{{}}\rangle
. $

The weakening of the linear electron-phonon interaction is expressed by the
coefficient $\exp (\vartheta _{q}^{{}})$ $\exp (\eta _{q}^{{}})$ . The value
of $\exp (\eta _{q}^{{}})$ is set by the diagonalization condition of the
anharmonic phonon Hamiltonian, Eq.\ref{10.4} according to the identity: 
\begin{equation}
e^{2\eta _{q}^{{}}}=\sqrt{\frac{1+\tanh \left( 2\eta _{q}^{{}}\right) }{%
1-\tanh \left( 2\eta _{q}^{{}}\right) }}  \label{eq.exp2eta}
\end{equation}

Therefore the squeezing effect related to the anharmonicity of the phonons
also reduces the electron-phonon interactions.

To conclude, let us check if the link that our model establishes between the
renormalization of the harmonic frequency from $\Omega _{q}^{{}}$ to $\omega
_{q}^{{}}$ and the reduction of the electron-phonon coupling strength is
consistent with the estimates of those quantities as given, e.g., in Refs. 
\cite{liu,Yildirim,choi}. The ratio $\Omega _{q}^{{}}$ /$\omega _{q}^{{}}$is
evaluated \ as $85\%$.\cite{liu}, $75\%$\cite{Yildirim} and $80\%$\cite{choi}%
. We assume that the value of $\exp (\vartheta _{q}^{{}})$ \ (Eq.\ref
{eq.thetaq}) , which we can not estimateat this stage, is not far from
unity. By taking from Ref.\cite{Yildirim} $x_{qq}^{{}}\sim 7.8$ eV$^{-1}$, $%
\hbar \Omega _{q}^{{}}=60$ meV, and assuming a dispersionless mode in Eq.\ref
{eqXq} we obtain, at zero temperature, where we can neglect $\langle \nu
_{q}^{{}}\rangle ,$ that $X_{q}^{{}}\sim 1.12$ . Then $\tanh \left( 2\eta
_{q}^{{}}\right) \sim -0.105$ yielding, from Eq.\ref{SQomega}, $\Omega
_{q}^{{}}/\omega _{q}^{{}}\sim 0.85$ \ and, from Eq.\ref{eq.exp2eta}, $\exp
\left( 2\eta _{q}^{{}}\right) =.90$. In Eliashberg's theory\cite{barisic} $%
\lambda \sim \langle |g^{2}|\rangle /\langle \omega ^{2}\rangle $, where $%
\langle ...\rangle $ are suitable averages. In our model then $\lambda
_{anhar}/\lambda _{har}\sim e^{2\eta _{q}^{{}}}/X_{q}^{2}\left[ 1-\tanh
^{2}\left( 2\eta _{q}^{{}}\right) \right] =0.65.$ which agrees with the
estimate \cite{choi} of a $30\%$ weakening of $\lambda _{anhar}$. From the
fact that \ \ $\tanh \left( 2\eta _{q}^{{}}\right) \sim -0.105$ we also
conclude that the large phonon softening found on passing from $AlB_{2}$ to $%
MgB_{2}$ \cite{Yildirim,bohnen,renker}\ is probably not due only to the
harmonic-to-anharmonic phonon change, because the squeezing-induced
softening effect in Eq.\ref{SQomega} is not strong enough. The
filling-dependent effect of $\exp (\vartheta _{q}^{{}})$ should also be
taken into account quantitatively before drawing more reliable conclusions
on this point.

\section{Conclusions.}

We have obtained a model Hamiltonian which should contain the essential
physics of $MgB_{2}.$\ 

Our starting point was a two-band Hamiltonian with anharmonic phonons and
both linear and quadratic electron-phonon interactions of the
bond-stretching type, as dictated by the results of LDA calculations \cite
{liu,Yildirim,kunc,choi,bohnen}. The final Hamiltonian followed by applying
a sequence of unitary transformations to both the electronic and phononic
terms. In particular, we have been able to go beyond the Hartree-Fock
approximation in treating the anharmonic effects. We have thus obtained an
effective Hamiltonian (Eqs.\ref{eq.47} and \ref{eq.49}) of a very simple
structure. The electronic part has two bands with a phonon-depending
hybridization, generated by the quadratic electron-phonon interaction, which
increases with temperature as observed\cite{dahm}. The phononic part has an
effective harmonic free-phonon term with a frequency which depends from the
band filling factors. Finally, the effective electron-phonon interaction is
reduced to a linear one, but with an amplitude also depending from the band
filling factors, which would result in a similar dependence of the phonon
linewidths\cite{postorino,renker}. One could ask if it might have been
possible to obtain the same results starting from a simpler Hamiltonian, as,
for instance, the one proposed in Ref.\cite{boeri}. We do not think so.
Indeed, all our results depend basically from taking into account the
quartic anharmonicity and the second order EPI, as can be checked by
considering the limits for $\eta _{q}\rightarrow 0$ and $\vartheta
_{q}\rightarrow 0$ of Eqs.\ref{eq.46},\ref{eq.47} and \ref{eq.49}. Namely,
if both $\eta _{q},\vartheta _{q}\rightarrow 0$, from Eq.\ref{eq.49}\ one
finds no reduction of the effective EPI. If there is no second-order EPI,
then $\Phi _{q}=0$\ follows, implying $\vartheta _{q}\rightarrow 0$, so that
there is no filling dependence of $\varpi _{q}$\ (Eq.\ref{eq.46}), no
phonon-number effect on the hybridization (Eq.\ref{eq.47}) and no filling
dependence of the phonon linewidths (Eq.\ref{eq.49}).

We have also shown that the numerical results of Ref.\cite{choi} about the
phononic features of the material, namely the renormalization of the
effective harmonic frequency $\omega _{q}^{{}}$ and the reduction of the
Eliashberg's $\lambda $, can be consistently interpreted as due to the
phonons accomodating themselves in a \ ''squeezed ''state. On the other
side, squeezing effects alone are not strong enough to account for the large 
$E_{2g}$ phonon softening on passing from $AlB_{2}$ to $MgB_{2}$ \cite
{Yildirim,bohnen,renker}.

\begin{acknowledgement}
We thank J. Spa\l ek, M. A. Gusm\~{a}o and J. R. Iglesias for stimulating
discussions.
\end{acknowledgement}

\end{document}